\documentclass[10pt, journal]{IEEEtran}
\IEEEoverridecommandlockouts
\usepackage{cite}
\usepackage{amsmath,amssymb,amsfonts}
\usepackage{algorithmic}
\usepackage{graphicx}
\usepackage{textcomp}
\usepackage{xcolor,float,tikz}
\def\BibTeX{{\rm B\kern-.05em{\sc i\kern-.025em b}\kern-.08em
    T\kern-.1667em\lower.7ex\hbox{E}\kern-.125emX}}
 \usepackage{tikz}  
 \usepackage{tkz-euclide}
\usepackage[utf8]{inputenc}
\usepackage{pgfplots} 
\usepackage{pgfgantt}
\usepackage{pdflscape}
\pgfplotsset{compat=newest}
\newlength\fheight
\newlength\fwidth    
\usepackage{mathtools}     
\usepackage{caption,lipsum}
\usepackage{subcaption}
\usepackage{balance}

\usepackage{amsthm}
\newtheorem*{remark}{Remark}
\begin{document}
\title{Situational Coverage and Rate Distribution Maps for 5G V2X Systems Using Ray-Tracing}

\author{Elyes~Balti,~\IEEEmembership{Member,~IEEE}
\thanks{Elyes Balti is with the Wireless Networking and Communications Group, Department of Electrical and Computer Engineering, The University
of Texas at Austin, Austin, TX 78712 USA (e-mails: ebalti@utexas.edu).}
}

\maketitle

\begin{abstract}
Millimeter wave (mmWave) is a practical solution to provide high data rate for vehicle-to-everything (V2X) communications. This enables the future autonomous vehicles to exchange big data with the base stations (BSs) such as the velocity, and the location to enhance the safety for the advanced driving assistance system (ADAS). To achieve this goal, we propose to develop a situational rate map to characterize the distribution of the rates achieved between the BSs and a uniform grid of vehicles. In this context, we consider a mmWave 5G cellular system with two physical structures which are the analog-only beamforming and hybrid precoding with limited feedback to investigate the rate distribution for single and multiuser scenarios. We will use the Ray-Tracing tool to construct the simulation environment and generate the channels between the BSs and the grid of vehicles. Finally, we will study the effects of the carrier frequency, the bandwidth, the BSs deployment, the blockage, the codebook, the physical architectures and the number of served users on the rate distribution maps. Moreover, we will present the rate statistics to evaluate the coverage of served users for certain services requirements and for various road shapes such as corners, intersections and straightways.
\end{abstract}


\section{Introduction}
The increase in demand for bandwidth has been growing exponentially over the last few decades due to the large number of subscribers and the number of communication devices \cite{j3}. Because of these factors, the users suffer from the spectrum scarcity as most of the available spectrum is totally assigned to the licensed users. In addition, spectrum sharing-sensing systems which consist of primary and secondary users also reach its bottleneck. Due to the massive number of users, the secondary users become unable to take advantages of the spectrum holes left by the primary users. Also, the primary users are not efficiently communicating since the licensed microwave spectrum becomes very limited. Furthermore, advanced Long Term Evolution (LTE) communications systems are in desperate need for high-speed wireless data rates in order to share big data, high definition (HD) videos, vehicular platooning/clouding, advanced driving assistance system (ADAS) for vehicular communications \cite{v1,v2,v3,c1}. Besides, self-driving vehicles requires a big amount of shared data with the other surrounding vehicles to help them exploring their environments and reduce the rates of accidents \cite{c4,n1,n2,c2,map}.

\begin{table*}[t]
    \caption{Features of DSRC Vs LTE for V2X \cite{va}.}\label{i1}
    \centering
    \begin{tabular}{|c|c|c|c|}
    \hline
    \textbf{Features} & \textbf{DSRC} & \textbf{D2D LTE-V2X} & \textbf{Cellular LTE-V2X}\\\hline    
    Channel width & 10 MHz & up to 20 MHz & up to 20 MHz  \\\hline
    Frequency band & 5.9 GHz & 5.9 GHz & 450 MHz - 3.8 GHz \\\hline 
    Bit rate & 3-27 Mbps & up to 44 Mbps & up to 75 Mbps  \\\hline
    Range & $\approx$ 100s m & $\approx$ 100s m & up to a few Km  \\\hline
    Spectral efficiency & 0.6 bps/Hz & 0.6 bps/Hz (typical) & 0.6 bps/Hz (typical)  \\\hline
    Coverage & ubiquitous & ubiquitous & ubiquitous  \\\hline
    Mobility support & high speed & high speed & high speed  \\\hline
    Latency & $\times$ ms & $\times$10$\times$100 ms & $\times$10 ms  \\\hline
    \end{tabular}
\end{table*}

\begin{table*}[t]
    \caption{Some use cases and network requirements for 5G V2X \cite{3gpp2,3gpp3,3gpp4,3gpp5,3gpp6}}\label{i2}
    \centering
    \begin{tabular}{|c|c|c|c|}
    \hline
    \textbf{Use case} & \textbf{Latency} & \textbf{Reliability} & \textbf{Data rate}\\\hline    
    Vehicle platooning & $\leq$ 25 ms & $\geq$ 90 $\%$ & low  \\\hline
    Remote driving & $\leq$ 5 ms & $\geq$ 99.99 $\%$ & $\geq$ 10 Mbps DL, $\geq$ 20 Mbps UL \\\hline 
    Collective perception of environment & $\leq$ 3 ms & $\geq$ 99 $\%$ & 1 Gbps for a single UE \\\hline
    Cooperative collision avoidance & $\leq$ 10 ms & $\geq$ 99.99 $\%$ & $\geq$ 10 Mbps \\\hline 
    Info sharing for level 2/3 aut. & $\leq$ 100 ms & $\geq$ 90 $\%$ & $\geq$ 0.5 Mbps \\\hline 
    Info sharing for level 4/5 aut. & $\leq$ 100 ms & $\geq$ 90 $\%$ & $\geq$ 50 Mbps \\\hline 
    Video data sharing for improved automated driving & $\leq$ 10 ms & $\geq$ 99.99 $\%$ & $\geq$ 100 - 700 Mbps \\\hline 
    \end{tabular}
\end{table*}

\subsection{Literature Review}
Autonomous connected vehicles have recently emerged as a new technology for future vehicular communications. Such networks consist of the vehicles and the roadside units (RSUs) or the base stations (BSs) to exchange the data related to the traffic flows, velocities, road conditions, safety warnings, and driving assistance, etc. Such information is very important for self-driving vehicles to avoid traffic congestions and potential collisions with cyclists and pedestrians. The issues may arise in complicated road intersections where the risk of potential accidents is severe. Typically, the onboard automotive sensors provide the required information for the vehicles to improve the safety awareness or the ADAS. Dedicated short-range communication (DSRC) has been implemented for vehicle-to-infrastructure (V2I) and vehicle-to-vehicle (V2V) to exchange basic sensor information within a range of 1 km and with a data rate from 2 to 6 Mbits/s \cite{va}. For LTE cellular systems, the maximum data rate is limited to 100 Mbits/s for high mobility while typically much lower rates are achieved \cite{va,lte}. TABLE \ref{i1} provides the features of DSRC and LTE for V2X. Thereby, conventional DSRC and LTE systems cannot support the Gbits/sec data rates required for the 5G autonomous vehicles. Millimeter Wave (mmWave) technology can be considered as a solution to support the big data rate required for the future generation vehicles since it provides higher bandwidth \cite{j4}. TABLE \ref{i2} illustrates some services requirements for 5G V2X. For V2V communications, mmWave has been extensively investigated for more than a decade ago \cite{decade}. Related works have tested mmWave vehicular networks for simpler transceivers and multiple-input-multiple-output (MIMO) large arrays systems \cite{surv,tassi}. In particular, mmWave V2I systems with analog architectures, i.e., the system supports only a single spatial stream, have been considered in urban scenario \cite{urban,inv,pos,prediction,c3} and in the road intersections \cite{spatially,cross} wherein the blockage, coverage, and rate were analyzed.

Furthermore, the ranges for mmWave signals propagation are short due to high pathloss degradation (high frequencies) and huge penetration loss in case of blockage. For example, in urban street canyon scenario wherein the buildings and blockers are dense, line-of-sight (LOS) V2V communications are not always available and even the non-line-of-sight (NLOS) received powers are very low to be detected and processed. To overcome these limitations, some related works proposed cooperative relaying communications to improve the received power and enhance the scalability of the networks \cite{j1,j2,c1}. Wireless systems assisted-relaying also overcome the blockage problem when the LOS links are not available by employing relays between the source and destination so that the transmission occurs in two time slots. Moreover, the sidelink (V2V) achievable rates are sometimes required to be estimated to decide which services can be supported. In a scenario wherein the sidelink is blocked but the uplink and downlink rates are available, then by assuming that the RSUs behave as decode-and-forward (DF) relays, the sidelink rate can be estimated by the minimum between the uplink and downlink rates \cite{j3}. Robustness of cooperative relaying has been intensively investigated in the literature. For example, the works \cite{c2,c3} examined the probability of outage and error for amplify-and-forward relaying wherein multiple relaying have been employed between the source and the destination and they showed that the performance gets much better by increasing the number of relays and hence exploiting the spatial diversity of relaying deployment. Other works focused on the effects of the co-channel interference and hardware impairments on the cooperative systems in terms of coverage and achievable rate. For example, related works considered cooperative systems operating at mmWave carrier frequency under feedback delay, hardware impairments and co-channel interference, and they derived tractable expressions for the coverage and rate \cite{j1,j2,j3}.

\subsection{Contribution}
In this paper, we propose to develop a situational map for 5G mmWave communications between the BSs and the vehicles as a function of their locations. Such rate map is useful to characterize the joint rate distribution in the environment and to analyze the set of rates achieved in different locations. In this context, we consider an urban street canyon scenario where the BSs are deployed at the street level and the grid of vehicles covers different road shapes. Moreover, analog architecture is assumed for the vehicles while we consider the cases of analog and hybrid architectures for the BSs. For the hybrid architecture, we consider the fully-connected and partially-connected structures for different scenarios and we will establish the trade-off between the achievable rate and the energy efficiency. For the system simulation, we consider the Wireless-Insite Ray-Tracing tool \cite{ray} to generate the channels between the BSs and the vehicles for different simulation environments. More specifically, this paper makes the following contributions
\begin{itemize}
    \item Using Ray-Tracing, we construct the environment including the buildings, roads, BSs deployment as well as the grid of vehicles.
    \item Extract the data from the output files such as the received powers, the phases, the angles of arrival and departure, and the delays in order to construct the MIMO channels between the BSs and the vehicles.
    \item Consider analog-only architecture for BSs and vehicles and study the single-user scenario with TDMA sharing.
    \item Consider the hybrid architecture for the BSs with limited feedback and study the multiuser scenario.
    \item Construct the beamformers and evaluate the achievable rates between the BSs and the vehicles.
    \item Construct the rate distribution map by associating each rate with the corresponding vehicle location.
    \item Study the effects of various parameters on the rate map such as the blockage, the number of antennas, the frequency, the bandwidth, the number of scheduled users, etc, for single and multiuser scenarios.
    \item Collect the rate statistics for various realizations and for different traffic intensity, BS deployment and road geometry like corners, intersections and straightways.
    \item Exploit these rate statistics to investigate the coverage for three services requirements such as collective perception of environment, video data sharing for improved automated driving and information sharing for level 2/3 automation.
\end{itemize}

\subsection{Structure}
The rest of the paper is organized as follows: Section II describes the analog architecture of the system while Section III presents the analysis of the hybrid architecture. Section IV illustrates the numerical results with the discussion while the concluding remarks and future directions are reported in Section V.

\section{Analog Architecture}
\subsection{System Model}
\begin{figure}[H]
    \centering
    \includegraphics[width=8.5cm]{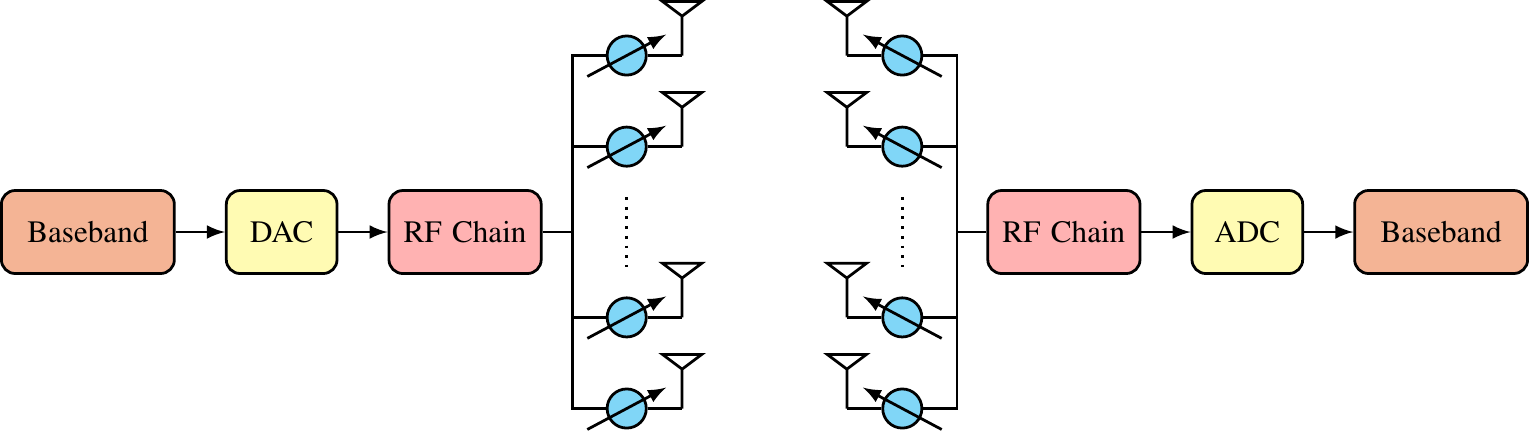}
    \caption{Analog architectures of the base station and the vehicle.}
    \label{analog}
\end{figure}

Fig.~\ref{analog} presents the physical structure of the BS and the vehicle. Such structure consists of a single RF chain to support only one spatial stream or vehicle. Thereby, the BS can serve only one vehicle at a time which is also called the single-user MIMO scenario. In addition, the array beam direction is managed only through the weights of the phase shifters as the amplitudes are constant for the analog architecture.
\subsection{Antennas Array Model}
In this work, we propose the uniform rectangular array (URA) with $N \times M$ elements where $N$ and $M$ are the vertical and horizontal dimensions of the array/subarray, respectively. This model also encompasses special cases of array structure such as the uniform linear array (ULA) or uniform square planar array (USPA). The array response of the URA is given by
\begin{equation}
\begin{split}
\textbf{a}(\phi,\theta) =&
 \frac{1}{\sqrt{NM}}\left[1,\ldots,e^{j\frac{2\pi}{\lambda}(d_hp\sin\phi\sin\theta+d_vq\cos\theta)},\ldots\right.\\&\left. ,e^{j\frac{2\pi}{\lambda}(d_h(M-1)\sin\phi\sin\theta+d_v(N-1)\cos\theta)}   \right]^{T}
\end{split}
\end{equation}
where $\lambda$ is the carrier wavelength, $d_h$ and $d_v$ are the antenna spacing in horizontal and vertical dimensions, respectively, $0 \leq p \leq M-1$, and $0 \leq q \leq N-1$ are the antenna indices in the 2D plane.

\subsection{Analog Beam Codebook}
Since 3D beamforming is assumed, we quantize the azimuthal $\phi$ and elevation $\theta$ angles along with an oversampling factor $\rho$ as $\phi_m, m = 1,\ldots,M$ and $\theta_n, n = 1,\ldots,N$. The $m$-th element $\nu_{m,k,\ell}$ of azimuthal beam $\nu_{k,\ell}$ and the $n$-th element $\delta_{n,\ell}$ of elevation beam $\delta_\ell$ are given by
\begin{equation}
    \nu_{m,k,\ell} = \frac{1}{\sqrt{M}}\exp\left(-j\frac{2\pi}{\lambda}(m-1)d_h\sin\phi_k\sin\theta_\ell\right)
\end{equation}
\begin{equation}
\delta_{n,\ell} = \frac{1}{\sqrt{N}}\exp\left(-j\frac{2\pi}{\lambda}(n-1)d_v\cos\theta_\ell\right)   
\end{equation}
where $\phi_k$ and $\theta_\ell$ are the $k$-th and $\ell$-th element of $\phi$ and $\theta$, respectively. Thereby, the $(k,\ell)$ entry of the codebook $\omega_{k,\ell}$ supporting 3D beamforming is given by the Kronecker product of the azimuthal and elevation array responses as 
\begin{equation}
\omega_{k,\ell} = \nu_{k,\ell} \otimes \delta_\ell.    
\end{equation}
\subsection{Beamforming Approach}
To find the best beam pair between the BS and the vehicle, we adopt the exhaustive beam search approach based on the designed codebooks at the BS and the vehicles. The BS and the $u$-th vehicle will select the analog precoder $\textbf{f}_u$ and combiner $\textbf{w}_u$ satisfying the following requirement
\begin{equation}
    (\textbf{w}_u,\textbf{f}_u) = \operatorname*{arg\,max}_{\textbf{w}\in \mathcal{W},~\textbf{f}\in \mathcal{F}} \|\textbf{w}^*\textbf{H}_u\textbf{f}\|
\end{equation}
where $(\cdot)^*$ is the Hermitian operator, $\textbf{H}_u$ is the channel between the BS and the $u$-th vehicle, $\mathcal{F}$, and $\mathcal{W}$ are the codebooks of the BS and the vehicles, respectively.
\begin{figure}[H]
\centering
\includegraphics[width=8.5cm]{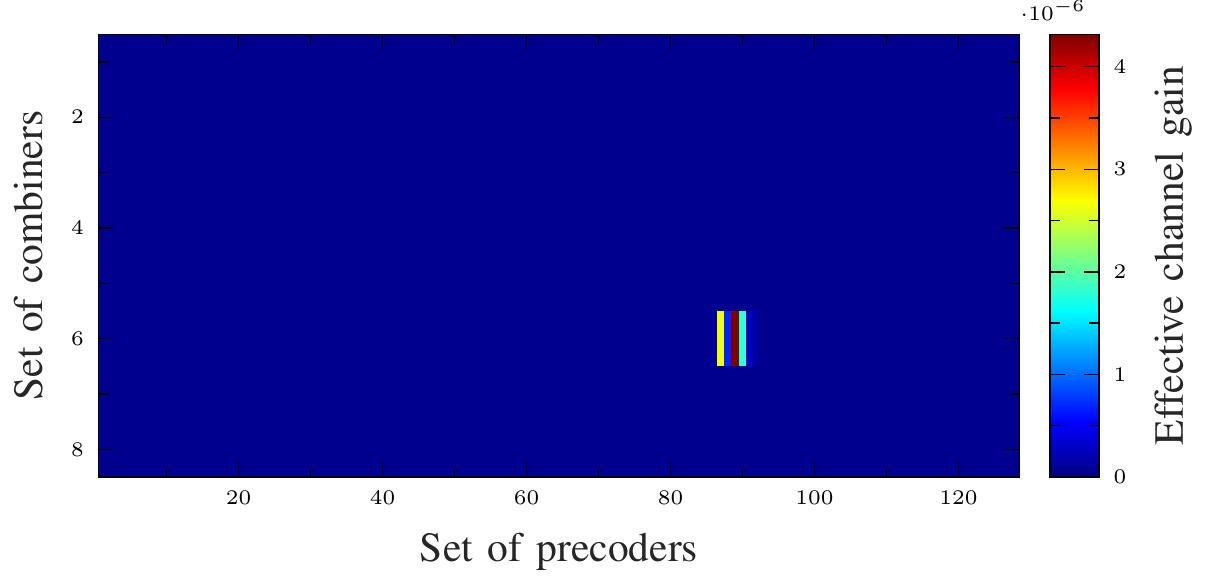}
    \caption{Example of beam search over codebooks of 128 precoders and 8 combiners.}
    \label{dft}
\end{figure}

Fig.~\ref{dft} illustrates the exhaustive beam search approach. The codebooks at the BS and the vehicle are with 128 and 8 codewords, respectively. The output of the exhaustive beam search algorithm yields the $6^{\text{th}}$ combiner and the $89^{\text{th}}$ precoder that maximize the effective channel gain.
\subsection{Achievable Rate Analysis}
After performing the single-user RF stage to construct the precoder and combiner, the achievable rate for the $u$-th vehicle is expressed as
\begin{equation}
R_u = B\log_2\left(1 + \frac{P|\textbf{w}_u^*\textbf{H}_u\textbf{f}_u|^2}{\sigma^2}\right)  
\end{equation}
where $B$ is the bandwidth, $P$ is the transmit power of the BS, and $\sigma^2$ is the noise power expressed as
\begin{equation}
\sigma^2[\text{dBm}] = -173.8 + 10 \log_{10}(B). 
\end{equation}

\section{Hybrid Precoding with Limited Feedback}
\subsection{System Model}
\begin{figure}[H]
    \centering
    \includegraphics[width=8.8cm]{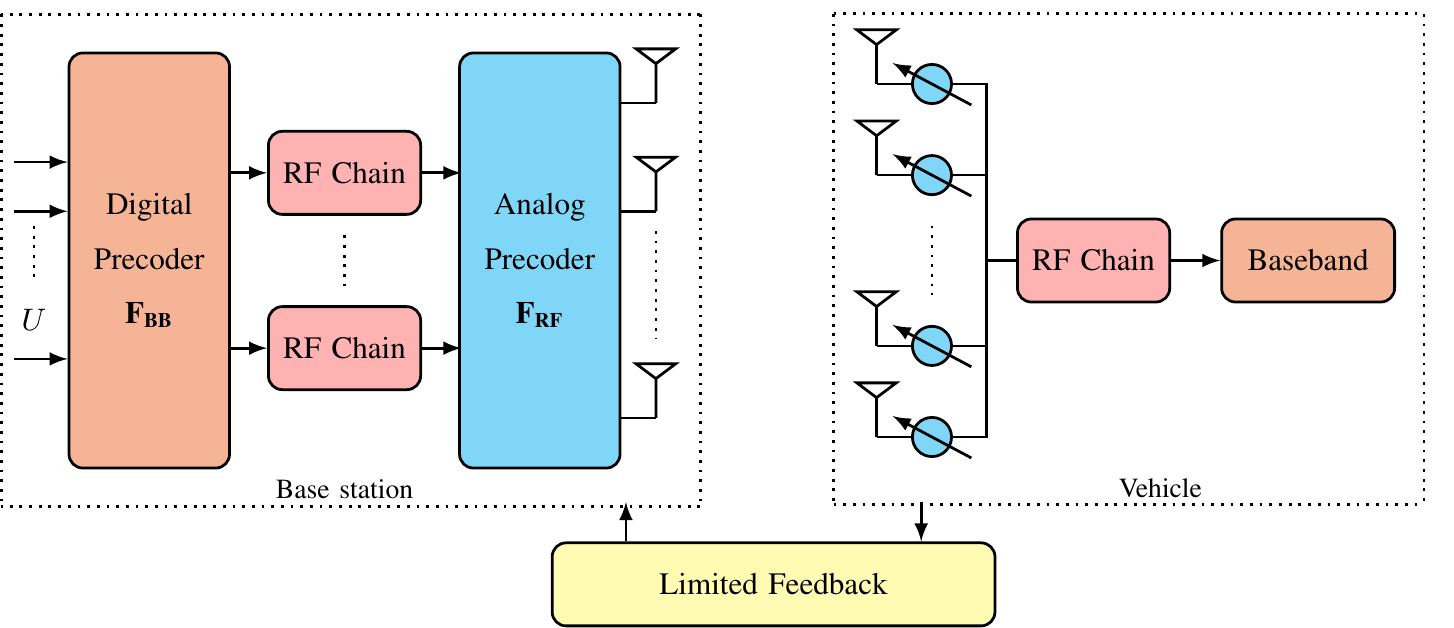}
    \caption{Physical architectures of the base station and the vehicle.}
    \label{hybrid}
\end{figure}

Fig.~\ref{hybrid} illustrates the physical architectures of the BS and the vehicle. The BS consists of hybrid digital/analog architecture supporting multiple spatial streams to serve different vehicles at a time. Once the best analog beamformers are selected in single-user RF stage, the $u$-th vehicle quantizes its effective channel ($\textbf{h}^*_u = \textbf{w}_u^*\textbf{H}_u\textbf{F}_{\text{RF}}$) and then feeds back the BS in order to construct the digital precoder. Note that $\textbf{F}_{\text{RF}}$ with size $N_t \times U$ is the analog precoder of the BS, $N_t$ is the number of the transmit antennas at the BS, and $U$ is the number of the served users at a time.

\subsection{Analog Precoding}
The analog beamformers are selected as in single-user RF stage between the BS and the served users. For the $u$-th vehicle, the best analog precoder $\textbf{f}_u$ and combiner $\textbf{w}_u$ are selected using exhaustive beam search detailed in subsection II.B. After repeating this process for all the selected users, the analog precoder of the BS is expressed as
\begin{equation}
    \textbf{F}_{\text{RF}} = [\textbf{f}_1,~\textbf{f}_2,\ldots,~\textbf{f}_U]
\end{equation}
where $\textbf{f}_u$ ($1\leq u \leq U$) is the selected analog precoder at the BS pointing to the $u$-th user.
\subsection{Digital Precoding}
This stage is a multiuser digital precoding design. For each user $u$, $u = 1,2,\ldots U$, the $u$-th vehicle quantizes its effective channel $\textbf{h}_u$ using a Random Vector Quantization (RVQ) codebook $\mathcal{H}$ and then feeds back $\hat{\textbf{h}}_u$ to the BS. The quantized channel $\hat{\textbf{h}}_u$ solves the following equation
\begin{equation}
\hat{\textbf{h}}_u = \operatorname*{arg\,max}_{\textbf{h}\in \mathcal{H}} \|\textbf{h}^*_u\textbf{h}\|.    
\end{equation}

For example, the Zero-Forcing digital precoding $\textbf{F}_{\text{BB}}$ is given by
\begin{equation}
\textbf{F}_{\text{BB}} = \textbf{H}^*\left(\textbf{H}\textbf{H}^*  \right)^{-1} 
\end{equation}
where $\textbf{H}$ is given by
\begin{equation}
\textbf{H} = [\hat{\textbf{h}}_1,\ldots,\hat{\textbf{h}}_U]^*.   
\end{equation}

\begin{remark}
Note that the number of served users should be reasonable as the channel codebook size grows exponentially with this number.
\end{remark}

The digital precoder of the $u$-th user has to be normalized as follows
\begin{equation}
\textbf{f}_u^{\text{BB}} = \frac{\textbf{f}_u^{\text{BB}}}{\|  \textbf{F}_{\text{RF}}\textbf{f}_u^{\text{BB}}\|_{F}},~u=1,2,\ldots,U.   
\end{equation}
\subsection{Hybrid Precoding Structures}
\subsubsection{Fully-Connected Structure}
For this structure, each RF chain is connected to all the phase shifters of the BS array. Although this structure achieves a higher rate as it provides more degree of freedom, it is not energy-efficient since a large amount of power is required for the connections between the RF chains and the phase shifters. The precoder pointed to the $u$-th user is a column vector of size $N_t \times 1$.
\subsubsection{Partially-Connected Structure}
For this structure, each RF chain is connected to a subarray of antennas of the BS array which reduces the hardware complexity in the RF domain. Although fully-connected outperforms the partially-connected structure in terms of achievable rate, the latter structure is well advocated for energy-efficient systems. Note that the analog precoder of the BS has the following structure
\begin{equation}
\textbf{F}_{\text{RF}} =
  \begin{pmatrix}
    \textbf{f}_1 & 0 & \dots & 0 \\
    0 & \textbf{f}_2 & \dots & 0 \\
    \vdots & \vdots & \ddots & \vdots \\
    0 & 0 & \dots & \textbf{f}_U
  \end{pmatrix}.
\end{equation}

The analog precoder pointed to the $u$-th vehicle ($\textbf{f}_u$) is retrieved by exhaustive beam search between the $u$-th subarray of the BS and the $u$-th vehicle. Such precoder is a column vector of size $N_{\text{sub}} \times 1$, and $N_{\text{sub}}$ is the number of antennas of the subarray.
\subsection{Performance Analysis}
\subsubsection{Achievable Rate Analysis}
As the BS serves multiple users at a time, each user $u$ is subject to the interference of the other users streams. Consequently, the achievable rate per user $u$ is given by
\begin{equation}
R_u = B\log_2\left(1 + \frac{\frac{P}{U}|\textbf{w}_u^*\textbf{H}_u\textbf{F}_{\text{RF}}\textbf{f}^{\text{BB}}_u|^2}{\frac{P}{U}\sum_{n \neq u}|\textbf{w}_u^*\textbf{H}_u\textbf{F}_{\text{RF}}\textbf{f}^{\text{BB}}_n|^2 + \sigma^2}\right).   
\end{equation}
\subsubsection{Energy Efficiency Analysis}
The energy efficiency per user $\eta_u$, expressed in bps/Watt or bits/Joule, is defined as the ratio between the achievable rate per user and the total power consumption. It is expressed as
\begin{equation}
\eta_u = \frac{R_u}{P_{\text{common}}+N_{\text{RF}}P_{\text{RF}}+N_tP_{\text{PA}} + N_{\text{PS}} P_{\text{PS}}}    
\end{equation}
where $N_{\text{RF}}$ is the number of RF chains, $P_{\text{common}}$ is the common power of the transmitter, $P_{\text{RF}}$ is the power of the RF chain, $P_{\text{PA}}$ is the power of the amplifier, and $P_{\text{PS}}$ is the power of the phase shifter. Note that $N_{\text{PS}}$ is given by
\begin{equation}
N_{\text{PS}} = \left\{
        \begin{array}{ll}
            N_tN_{\text{RF}} & \quad \text{Fully-connected} \\
            N_t & \quad \text{Partially-connected}
        \end{array}
    \right. .
\end{equation}

\section{Numerical Results}
\begin{figure*}[t]
\centering
\includegraphics[width=15cm]{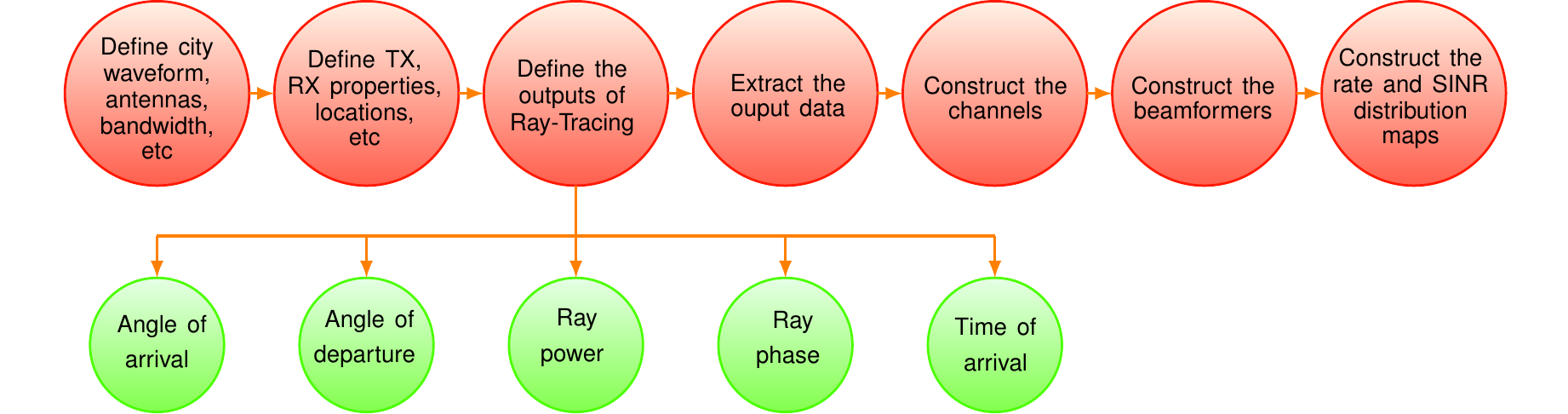}
    \caption{Simulation diagram.}
     \label{diagram}
\end{figure*}
In this work, we adopt the ray-based channel model between the BS and the $u$-th user which is given by
    \begin{equation}\label{channel}
        \textbf{H}_u = \sum_{\ell=0}^{L-1}\sqrt{p_\ell} e^{j\phi_\ell}e^{j2\pi f_c \tau_\ell}\textbf{a}_r(\varphi^r_\ell,\theta^r_\ell)\textbf{a}_t^*(\varphi^t_\ell,\theta^t_\ell)
    \end{equation}
where $L$ is the number of rays between the BS and the $u$-th user, $p_\ell$, and $\phi_\ell$ are the power and phase of the $\ell$-th ray, $f_c$ is the carrier frequency, $\tau_\ell$ is the delay of the $\ell$-th ray, $\varphi^r_\ell$ and $\theta^r_\ell$ are the azimuthal and elevation angles of arrival, respectively, while $\varphi^t_\ell$ and $\theta^t_\ell$ are the azimuthal and elevation angles of departure. $\textbf{a}_t(\cdot,\cdot)$ and $\textbf{a}_r(\cdot,\cdot)$ are the array steering and response vectors at the BS and the $u$-th user, respectively. 

We refer to the Wireless Insite Ray-Tracing tool \footnote[1]{Wireless Insite Ray-Tracing is a software tool developed by Remcom. This tool is based on generating an electromagnetic (EM) waves from the transmitter to the receiver and it predicts the channel behavior by accurately accounting for the EM radio propagation, reflection, diffraction, etc in different environments such as indoor, outdoor urban and rural areas, etc. More information about this tool are found in \cite{ray}} to construct the simulation environment and generate the channels between the BS and the grid of vehicles. In particular, Fig.~\ref{diagram} illustrates the diagram of the simulation setup where the required input data for the channel (\ref{channel}) are defined. 

To develop a smoother rate map, we consider an interpolation of 8 points between two successive vehicles. Unless otherwise stated, TABLES \ref{cityross} and \ref{sim} summarize the main simulation parameters of the terrain and the system.

\begin{table}[H]
    \caption{City Properties}\label{cityross}
    \centering
    \begin{tabular}{|c|c|c|c|}
    \hline
    \textbf{Material} & \textbf{Permittivity} & \textbf{Conductivity} & \textbf{Thickness}\\\hline    
    Building (concrete) & 15 & 0.015 S/m & 0.3 m  \\\hline
    Terrain (wet earth) & 25 & 0.02 S/m & 0 m \\\hline 
    \end{tabular}
\end{table}
\begin{table}[H]
    \caption{Simulation Parameters}
    \centering\label{sim}
    \begin{tabular}{|c|c|}
    \hline
    \textbf{Parameters} & \textbf{Value} \\\hline
    Number of BS antennas     & 256  \\\hline
    Number of vehicle antennas    & 4 \\\hline
    Input power & 10 dBm \\\hline
    Carrier frequency & 28 GHz \\\hline
    Bandwidth & 850 MHz \\\hline
    Vehicles interspace & 5 m \\\hline
    \end{tabular}
 \label{sim}
\end{table}

\subsection{Analog Precoding and Single-User MIMO}
For this scenario, as the BS serves only one user, we will map all the rates achieved between the BS and each user to the corresponding location of the latter in the grid. In this part, we will focus on the impacts of the blockage and the codebook on the rate map.

\begin{figure}[H]
        \centering
        \begin{subfigure}[b]{0.2\textwidth}
\centering
\includegraphics[width=3.7cm]{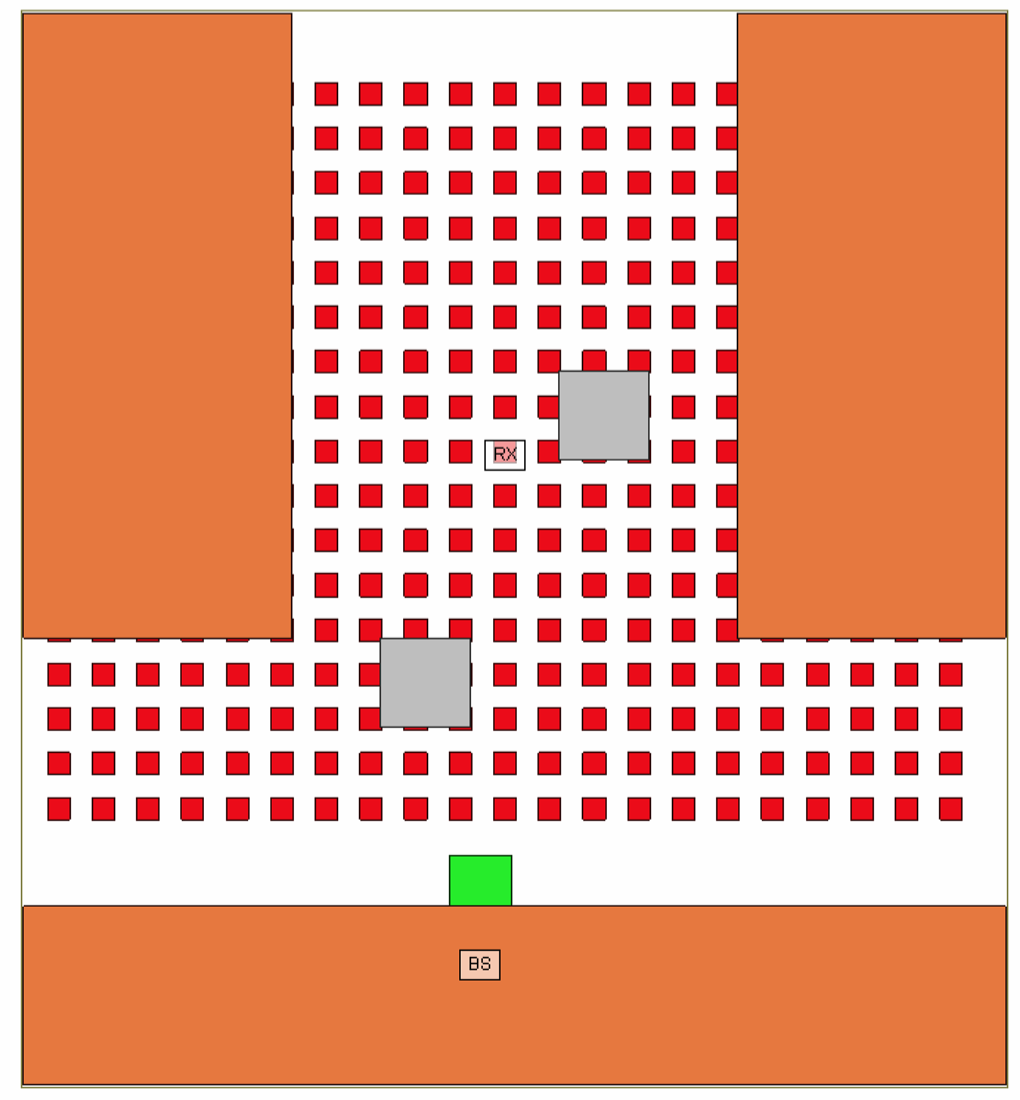}
  \caption{Top view of the terrain.}
    \label{a1}   
        \end{subfigure}\hspace*{0cm}
        \begin{subfigure}[b]{0.2\textwidth}  
\centering
\includegraphics{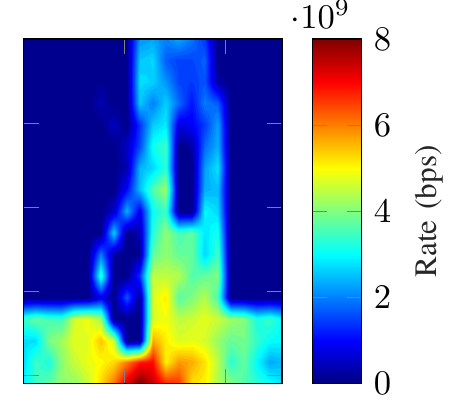}
  \caption{Regular codebook.}
    \label{a2}
        \end{subfigure}
        \vskip\baselineskip
        \begin{subfigure}[b]{0.2\textwidth}   
\centering
\includegraphics{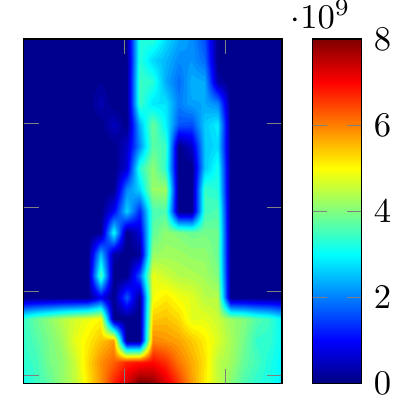}
  \caption{Oversampling = 4.}
    \label{a3}
        \end{subfigure}
        \quad \hspace*{-0.45cm}
        \begin{subfigure}[b]{0.2\textwidth}   
\centering
\includegraphics{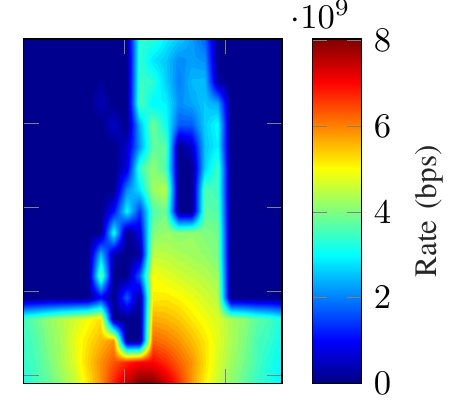}
  \caption{Oversampling = 16.}
    \label{a4}
        \end{subfigure}
        \caption[ ]
        {\small Rate maps generated for a blockage scenario with regular and oversampled codebook.} 
        \label{pic1}
    \end{figure}

Fig.~\ref{pic1} provides the rate maps performance generated for different oversampling factors of the codebook. The grid of receivers consists of two blocking metal trucks of 5 m height. In Fig.~\ref{a2}, we observe the corresponding locations of the blockage in the rate map where the NLOS users achieve lower rates. It is important to note that there are some users, even though blocked by the trucks, they still achieve non zero rates. This is explained by the fact that the signals reach those NLOS users by reflections from the buildings located at the roadsides. We also observe that by increasing the oversampling factor of the codebook, the users achieve better rates shown by Fig.~\ref{a3}, and Fig.~\ref{a4}. In fact, when the codebook is more refined, the BS array becomes able to scan and cover more directions in the grid of the receivers resulting in further rates improvement.

In the next simulation scenario, we will study the impacts of number of antennas, the carrier frequency, and the bandwidth on the rate map. In particular, we will consider the Rosslyn city as a simulation environment and we deploy the BSs at different locations in the city. For a given location, we evaluate the rates achieved between the users and the serving BS. Next, we repeat this scenario for different BSs locations and then we combine the results to generate the rate map. Specifically, we will consider the values of the carrier frequency and the corresponding bandwidth defined by the 5G New Radio (NR) standards and the ones being discussed for future deployments in US \cite{ni,anum}. Note that the buildings and the terrain properties have the same values given by TABLE \ref{cityross}.

\begin{figure}[H]
    \centering
    \includegraphics[width=\linewidth]{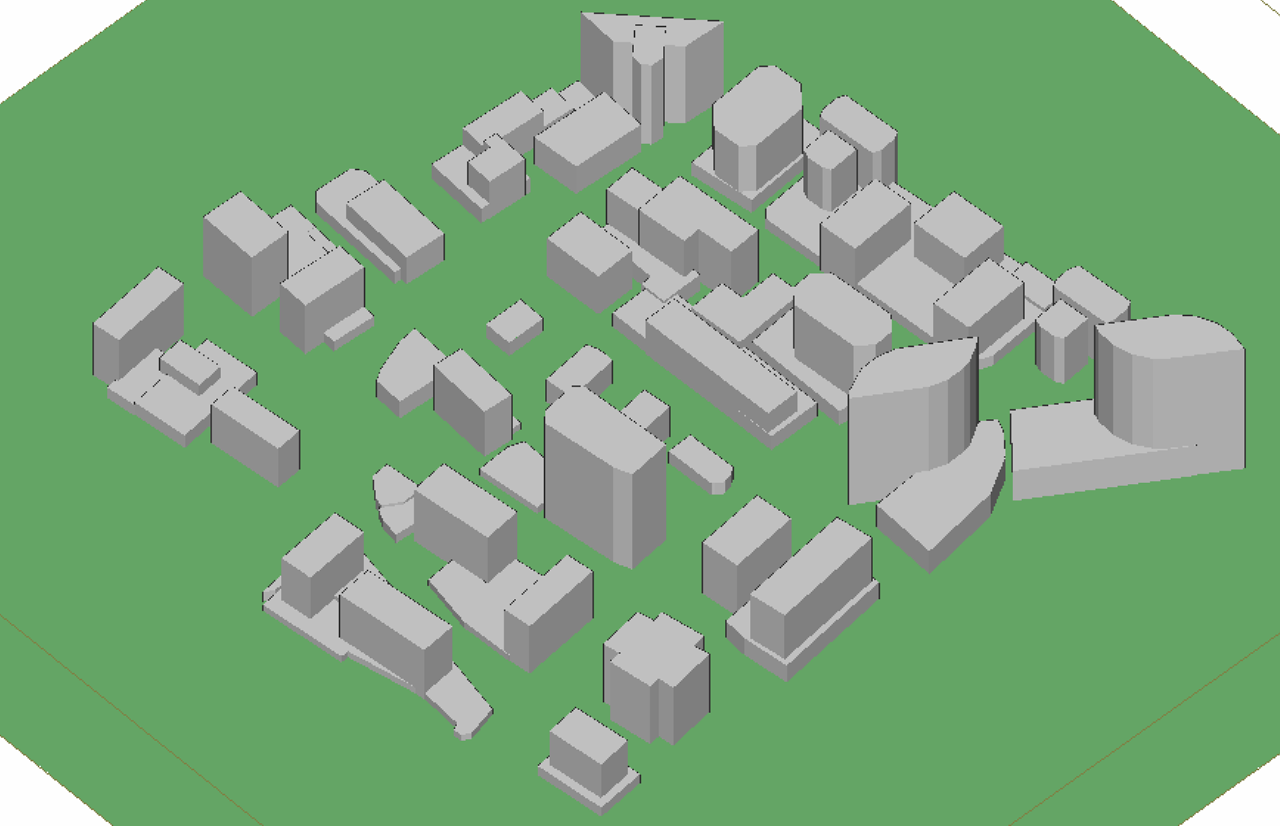}
    \caption{3D top view of Rosslyn city after segmentation.}
    \label{fig:my_label}
\end{figure}

 \begin{figure}[H]
        \centering
        \begin{subfigure}[b]{0.2\textwidth}
\centering
\includegraphics[width=3.7cm]{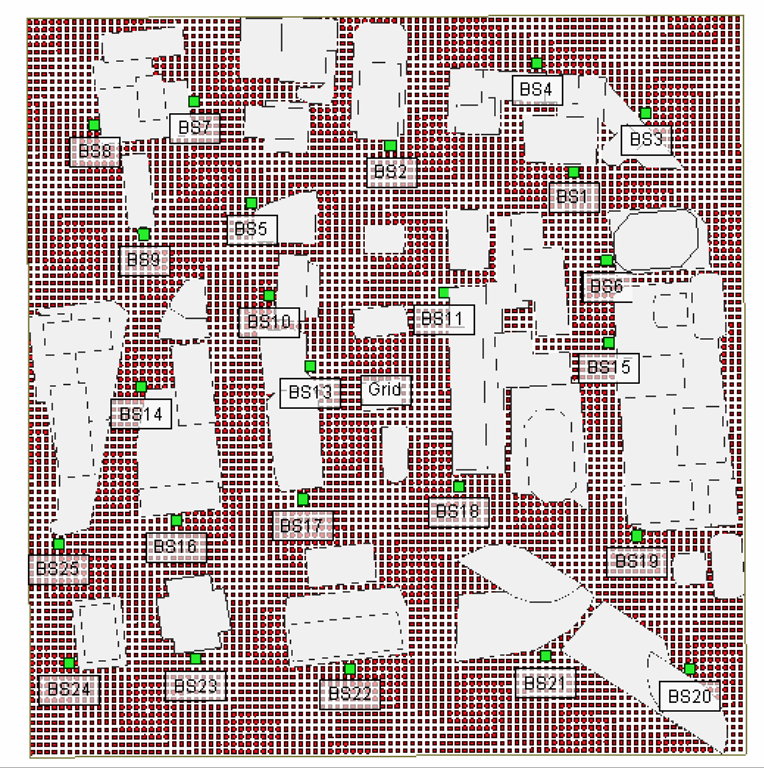}
  \caption{Top view of the Rosslyn city: 24 BSs are deployed.}
    \label{b1}   
        \end{subfigure}\hspace*{0.4cm}
        \begin{subfigure}[b]{0.2\textwidth}  
\centering
\includegraphics{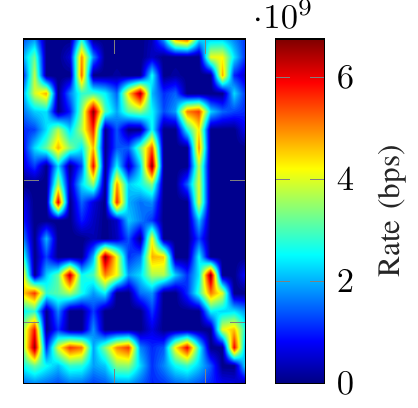}
  \caption{$f_c = 39$ GHz, $B = 1.6$ GHz, $N_t = 64$, $N_r = 4$.}
    \label{b2}
        \end{subfigure}
        \vskip\baselineskip
        \begin{subfigure}[b]{0.2\textwidth}   
\centering
\includegraphics{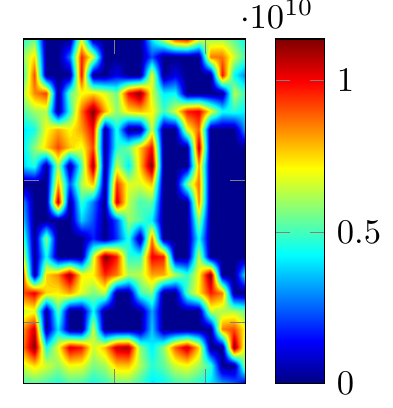}
  \caption{
 $f_c = 39$ GHz, $B = 1.6$ GHz, $N_t = 128$, $N_r = 16$.}
    \label{b3}
        \end{subfigure}
        \quad \hspace*{-0.5cm}
        \begin{subfigure}[b]{0.2\textwidth}   
\centering
\includegraphics{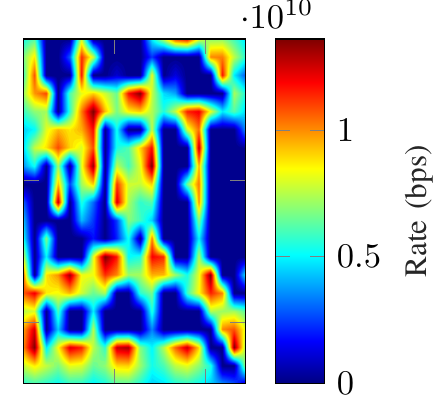}
  \caption{
  $f_c = 73$ GHz, $B = 2$ GHz, $N_t = 128$, $N_r = 16$.}
    \label{b4}
        \end{subfigure}
        \caption[ ]
        {\small Rate maps generated for the Rosslyn city. The BSs are shown by green color and the vehicles grid is shown by red color where the rates achieved between the BSs and the users are mapped accordingly.} 
        \label{pic2}
    \end{figure}

Fig.~\ref{b2} illustrates the rate distribution over the city for 24 BSs deployed at different locations. With such BSs deployment, most of the users are covered and they achieve non-zero rates. However, the rate distribution changes for LOS and NLOS users as the pathloss exponent and the blockage density are severe in this urban street canyon scenario. Fig.~\ref{b3} provides the rate map developed for the whole city but with higher number of antennas. Basically, the users achieve higher rates compared to the previous scenario. This is explained by the high beamforming gain achieved by the massive number of antennas and exhaustive beam search yielding higher achievable rates. Fig.~\ref{b4} presents the rate map but with higher carrier frequency and bandwidth compared to the scenario in Fig.~\ref{b3}. The maps show that the peak achievable rates increase with the bandwidth (12 Gbps for $B$ = 1.6 GHz, 15 Gbps for $B$ = 2 GHz). Given that the pathloss becomes more severe with the carrier frequency, the number of antennas at the BS is 256 which is sufficiently enough to provide a huge array gain to compensate for the pathloss degradation.

\subsection{Hybrid Precoding and Multiuser MIMO}
 \begin{figure}[H]
        \centering
        \begin{subfigure}[b]{0.2\textwidth}
\centering
\includegraphics[width=3.7cm,height=3cm]{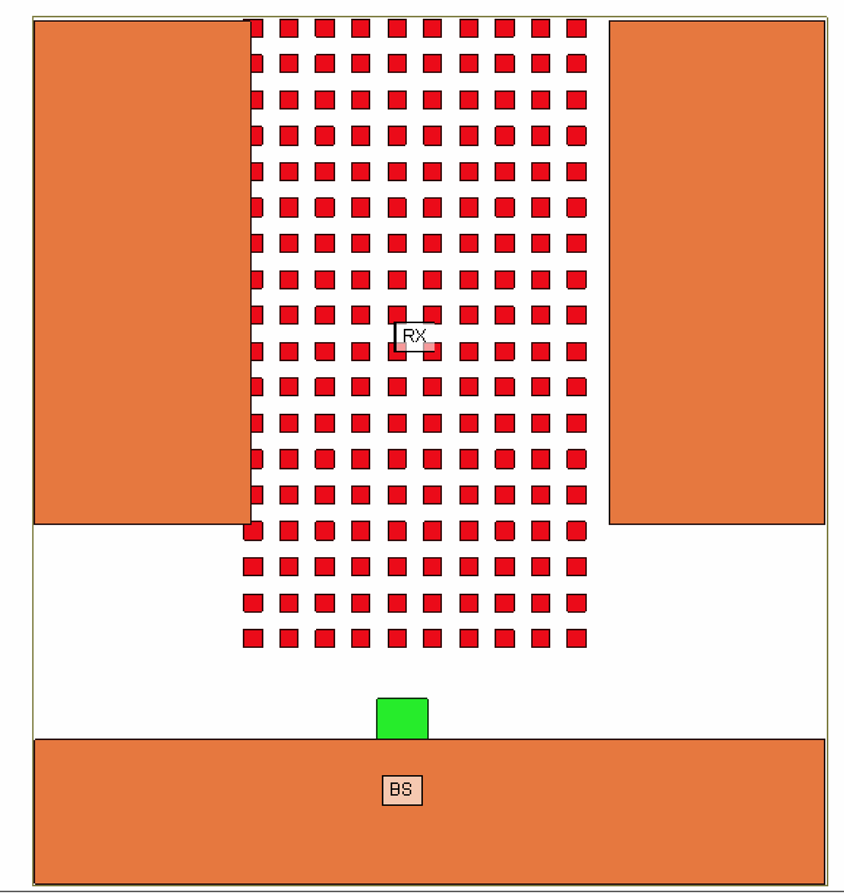}
  \caption{City layout.}
    \label{c1}   
        \end{subfigure}\hspace*{0cm}
        \begin{subfigure}[b]{0.2\textwidth}  
\centering
\includegraphics{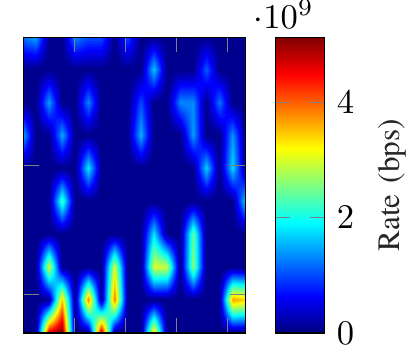}
  \caption{Perfect CSIT.}
    \label{c2}
        \end{subfigure}
        \vskip\baselineskip
        \begin{subfigure}[b]{0.2\textwidth}   
\centering
\includegraphics{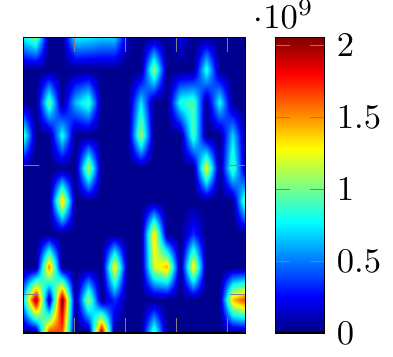}
  \caption{8 bits quantization.}
    \label{c3}
        \end{subfigure}
        \quad \hspace*{-0.3cm}
        \begin{subfigure}[b]{0.2\textwidth}   
\centering\includegraphics{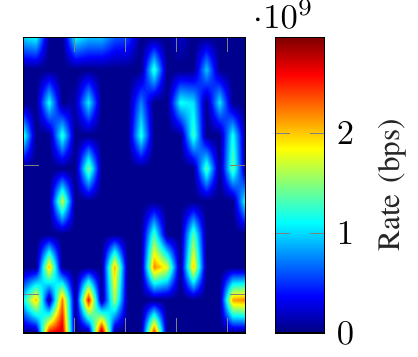}
  \caption{13 bits quantization.}
    \label{c4}
        \end{subfigure}
        \caption[ ]
        {\small Rate maps performance evaluated for an average number of realizations $N =10$. Fully-connected structure is assumed.} 
        \label{pic2}
    \end{figure}
    
In this simulation scenario, the BS randomly picks $U$ users from the cell and serve them at a time. Then, we repeat this selection process $N$ times to generate the situational rate map evaluated on average. Also we assume that $N_{\text{RF}} = U$. The total number of the users in the cell is 180 and the number of served users at a time is $U = 5$ users. Unlike the single-user MIMO case, we will map the rates only to the served users for this simulation scenario. The powers of the RF components are $P_{\text{common}} = 10$ W, $P_{\text{RF}} = P_{\text{PA}} = 100$ mW, and $P_{\text{PS}} = 10$ mW \cite{energy}. In this part, we will study the effects of the effective channels quantization, the number of served users, and the digital precoding on the rate map. Also we will compare the fully-connected and partially-connected structures in terms of rate and energy efficiency maps.

Fig.~\ref{pic2} illustrates the impact of the number of quantization bits on the rate map. When the effective channel is poorly quantized (8 bits), the loss incurred by multiuser interference is larger. However, the rate loss per user decreases when the number of quantization bits increases and the relative rate map becomes closer to the perfect CSIT rate map. 
\begin{figure}[H]
\begin{subfigure}[b]{0.2\textwidth}   
\centering
\includegraphics{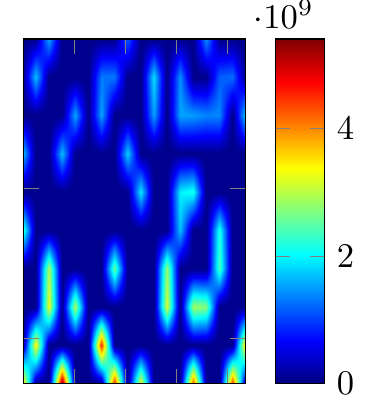}
  \caption{Fully-connected.}
    \label{d1}
        \end{subfigure}
        \quad 
        \begin{subfigure}[b]{0.2\textwidth}   
\centering\includegraphics{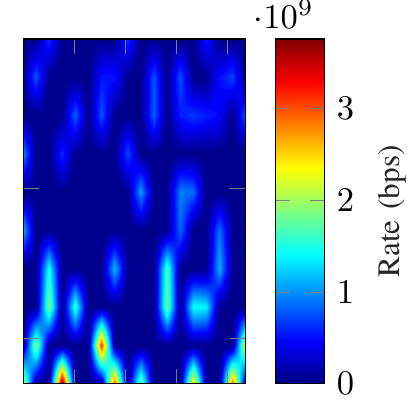}
  \caption{Partially-connected.}
    \label{d2}
        \end{subfigure}
        \caption[ ]
        {\small Rate maps performance evaluated for an average number of realizations $N =15$, perfect CSIT, and 4 served users. For the partially-connected structure, the BS has 4 subarrays each with 64 antennas to serve 4 users at a time. The total number of transmit antennas at the BS is 256 for both structures.} 
        \label{pic3}
\end{figure}

\begin{figure}[H]
\begin{subfigure}[b]{0.2\textwidth}   
\centering
\includegraphics{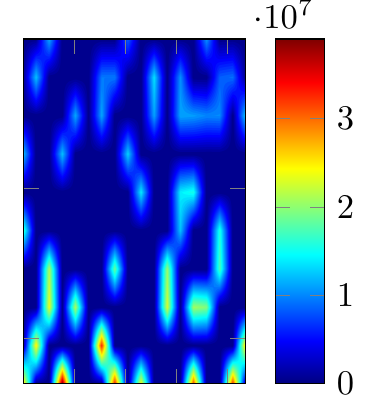}
  \caption{Fully-connected.}
    \label{d1}
        \end{subfigure}
        \quad 
        \begin{subfigure}[b]{0.2\textwidth}   
\centering
\includegraphics{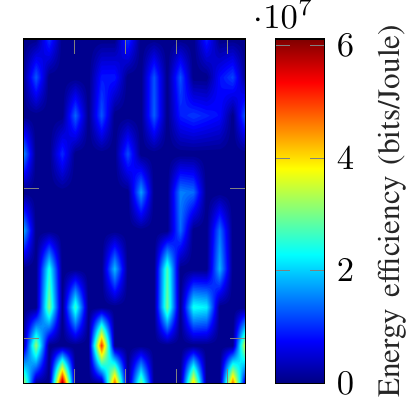}
  \caption{Partially-connected.}
    \label{d2}
        \end{subfigure}
        \caption[ ]
        {\small Energy efficiency maps performance evaluated for an average number of realizations $N =15$, perfect CSIT, and 4 served users. For the partially-connected structure, the BS has 4 sub-arrays each with 64 antennas to serve 4 users at a time. The total number of transmit antennas at the BS is 256 for both structures.} 
        \label{pic4}
\end{figure}   

Fig.~\ref{pic3} provides a comparison between two rate maps generated for fully-connected and partially-connected structures. As expected, the fully-connected structure achieves higher rates as much more degree of freedom are provided compared to the partially-connected structure.

In Fig.~\ref{pic4}, we observe that partially-connected structure outperforms the fully-connected structure in terms of power consumption. The main difference between the two structures is the number of connections between each RF chain and the phase shifters. In fact, the fully-connected structure requires $U$ times phase shifters power consumption more than the partially-connected structure. Thereby, the later structure achieves better energy efficiency.     

\begin{figure}[H]
\begin{subfigure}[b]{0.2\textwidth}   
\centering
\includegraphics{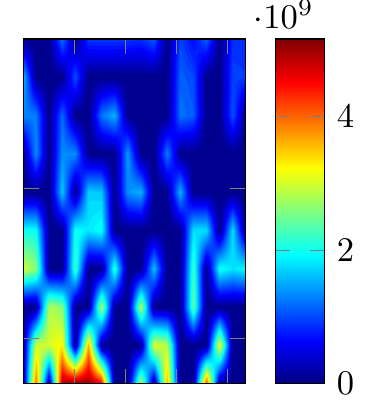}
  \caption{$U = 6$ served users.}
    \label{d1}
        \end{subfigure}
        \quad 
        \begin{subfigure}[b]{0.2\textwidth}   
\centering
\includegraphics{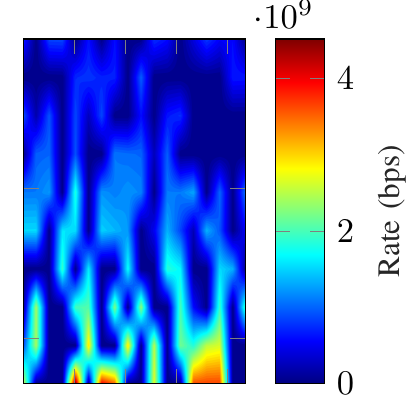}
  \caption{$U = 10$ served users.}
    \label{d2}
        \end{subfigure}
        \caption[ ]
        {\small Rate maps performance evaluated for an average number of realizations $N =15$, perfect CSIT, and fully-connected structure.} 
        \label{pic5}
\end{figure}  

\begin{figure}[H]
\begin{subfigure}[b]{0.2\textwidth}   
\centering
\includegraphics{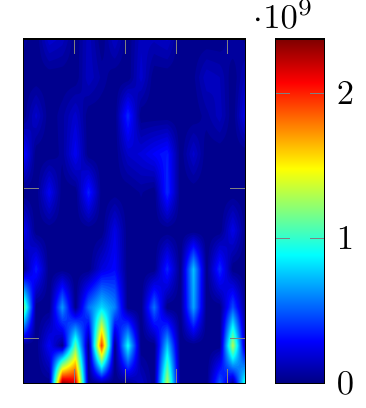}
  \caption{Zero-Forcing digital precoding.}
    \label{d1}
        \end{subfigure}
        \quad 
        \begin{subfigure}[b]{0.2\textwidth}   
\centering
\includegraphics{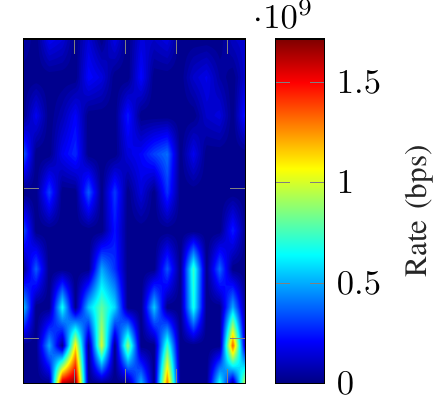}
  \caption{Without digital precoding ($\boldsymbol{F}_{\text{BB}} = \boldsymbol{I}_U$).}
    \label{d2}
        \end{subfigure}
        \caption[ ]
        {\small Rate maps performance evaluated for an average number of realizations $N =10$, perfect CSIT, and partially-connected structure. Each subarray having 32 elements serves one user at a time. The total number of transmit antennas at the BS is 256 and the number of served users is 8 on average.} 
        \label{pic6}
\end{figure}  

Fig.~\ref{pic5} illustrates the impact of the number of served users on the rate map. When the BS serves larger number of users, the average rate per user decreases and vice versa. This is explained by the fact that by increasing the number of spatial streams, each user becomes more vulnerable to the interference that grows with the number of users simultaneously served.

Fig.~\ref{pic6} presents the impact of the digital precoding on the rate map. We observe that the performance improves when considering the Zero-Forcing precoding compared to the scenario without digital precoding. In fact, the interference between users is well managed under Zero-Forcing precoding while it is more severe without digital processing.

\subsection{Rate Statistics and Coverage Area}
In this subsection, we will provide the statistics related to some V2X services such as info sharing for level 2/3 automation, video data sharing and collective perceptions of environment, which require target rates of 50 Mbps, 500 Mbps and 1 Gbps, respectively. We will focus on the reliability of each service in terms of rate and coverage in corners, intersections and straightways. We extract these road geometries from the Rosslyn city and we consider single-user scenario with TDMA sharing.

Once a transmission strategy is specified, the corresponding outage probability for rate $R$ (bits/s/Hz) is then
\begin{equation}
    P_{\textsf{out}}(\textsf{\scriptsize{SNR}}, R) = \mathbb{P}[\mathcal{I}(\textsf{\scriptsize{SNR}})<R].
\end{equation}

With convenient powerful channel codes, the probability of error when there is no outage is very small and hence the outage probability is an accurate approximation for the actual block error probability. As justified in the literature, modern radio systems such as UMTS and LTE operate at a target error probability. Therefore, the primary performance metric is the maximum rate\footnote[2]{In this work, we define the notion of rate with outage as the average data rate that is correctly received/decoded at the receiver which is equivalent to the throughput. In other standards in the literature, the rate with outage is assimilated with the transmit data rate. The only difference is if we consider rate with outage as the throughput, we account for the probability of bursts (outage) and we multiply by the term (1-$\epsilon$), while for the transmit data rate, the term (1-$\epsilon$) is not accounted anymore.}, at each {\textsf{\scriptsize{SNR}}}, such that this threshold is not overtaken, i.e.,
\begin{equation}
R_\epsilon(\textsf{\scriptsize{SNR}}) = \max_{\zeta}\left\{ \zeta: P_{\textsf{out}}(\textsf{\scriptsize{SNR}}, \zeta) \leq \epsilon \right\}    
\end{equation}
where $\epsilon$ is the target.
\subsubsection{Corners}
In this part, we will present the rate statistics averaged over 5 corners and 5 traffic realizations. Results are provided in terms of average rate, standard deviation and coverage in the following TABLES.
\begin{table}[H]
\caption{Results of averaging among 5 corners and 5 traffic realizations.}
    \centering
    \begin{tabular}{|c|c|c|c|c|}
    \hline
    \textbf{BS deployment}&\multicolumn{2}{|c|}{sparse}& \multicolumn{2}{|c|}{dense}\\ \hline
    \textbf{Traffic intensity} & light& high& light& high \\ \hline
    \textbf{Average rate (Gbps)} & 0.830& 0.802 & 1.250 & 1.230 \\ \hline
    \textbf{Standard deviation} &  0.027 & 0.087 & 0.089 & 0.091 \\ \hline
    \end{tabular}
 \end{table} 
 
  \begin{table}[H]
    \caption{Coverage percentage vs target rate for corners scenarios.}
    \centering
    \begin{tabular}{|c|c|c|c|}
    \hline
    \textbf{Target rate} & 1 Gbps & 500 Mbps & 50 Mbps\\\hline    
    \textbf{Coverage [$\%$]} & 75.70 & 93.20 & 97.80  \\\hline
    \end{tabular}
\end{table}

We observe that the traffic intensity has a low impact on the average rate while the BS deloyment density has a substantial effect on the rate. This is explained by the fact that densifying the BS deployment offers higher chances for the user to connect to the best serving BS which may alleviate the effects of the blockage in such complicated environment. For example, we notice that for the light traffic intensity, the rate increases from 0.830 to 1.250 Gbps which corresponds to an increase of around $33.6\%$ achieved by densifying the BS deployment. Moreover, the coverage is still acceptable as around $75.70\%$ of served users can support the collective perception of environment service which requires a rate requirement of 1 Gbps. More served users up to 93.20 $\%$ and 97.80$\%$ can support the other two services as the target rate requirements are relatively lower compared to the first service. 
 
\subsubsection{Intersections}
We move to evaluating the coverage and rate performance for the intersections environment. Statistics are collected for 5 different intersections and averaged over 5 traffic realizations.
\begin{table}[H]
\caption{Results of averaging among 5 intersections and 5 traffic realizations.}
    \centering\label{int1}
    \begin{tabular}{|c|c|c|c|c|}
    \hline
    \textbf{BS deployment}&\multicolumn{2}{|c|}{sparse}& \multicolumn{2}{|c|}{dense}\\ \hline
    \textbf{Traffic intensity} & light& high & light& high \\ \hline
    \textbf{Average rate (Gbps)} & 1.511& 1.495 & 1.834 & 1.805 \\ \hline
    \textbf{Standard deviation} & 0.119 & 0.102 & 0.101 & 0.129 \\ \hline
    \end{tabular}
 \end{table}    
 
 \begin{table}[H]
    \caption{Coverage percentage vs target rate for intersections scenarios.}\label{city}
    \centering\label{int2}
    \begin{tabular}{|c|c|c|c|}
    \hline
    \textbf{Target rate} & 1 Gbps & 500 Mbps & 50 Mbps\\\hline    
    \textbf{Coverage [$\%$]} & 82.48 & 94.08 & 99.44  \\\hline
    \end{tabular}
\end{table}

According to TABLES \ref{int1} and \ref{int2}, the traffic intensity still has negiligble effect on the average rate while the densification of BS deployment achieves around $22.5\%$ increase in rate. We also observe that the coverage for the three services becomes better compared to the corners scenarios. For example, the percentage of served users in coverage for the collective perception service increases from $75.5\%$ to 82.48$\%$ for corners and intersections, respectively. This improvement for coverage is expected since the intersections environments are characterized by spacious radio environments and less blockages density compared to the corners scenarios.

\subsubsection{Straightways}
In this section, we will discuss the rate statistics collected for 5 straightways environments and averaged over 5 realizations.
\begin{table}[H]
\caption{Results of averaging among 5 straightways and 5 traffic realizations.}
    \centering\label{st1}
    \begin{tabular}{|c|c|c|c|c|}
    \hline
    \textbf{BS deployment}&\multicolumn{2}{|c|}{sparse}& \multicolumn{2}{|c|}{dense}\\ \hline
    \textbf{Traffic intensity} & light& high & light& high \\ \hline
    \textbf{Average rate (Gbps)} & 1.859 & 1.847 & 2.400 & 2.380 \\ \hline
    \textbf{Standard deviation} & 0.067 & 0.125 & 0.084 & 0.103 \\ \hline
    \end{tabular}
 \end{table}    
 
 \begin{table}[H]
    \caption{Coverage percentage vs target rate for straightways scenarios.}
    \centering\label{st2}
    \begin{tabular}{|c|c|c|c|}
    \hline
    \textbf{Target rate} & 1 Gbps & 500 Mbps & 50 Mbps\\\hline    
    \textbf{Coverage [$\%$]} & 96.30 & 98.20 & 99.98  \\\hline
    \end{tabular}
\end{table}

Referring to TABLE \ref{st1}, the traffic intensity and the BS deployment still have similar effects on the average rate compared to the corners and intersections scenarios. From TABLE \ref{st2}, the percentage of users in coverage for the services with target rates 500 Mbps and 50 Mbps are roughly similar to the coverages for corners and intersections scenarios. However, the coverage relative to the collective perception service considerably enhances for the straightways environments as it jumps from 82.48$\%$ (intersections) to 96.30$\%$. This result is expected since the straightways are the less complicated scenario and most likely the served users are in LOS connectivity with the serving BS.
\subsubsection{BS Deployment: Pseudorandom vs Smart}
In the following TABLE \ref{bs}, we provide the spatial coverage results for pseudorandom (before) and smart (after) deployment. The results are collected for corners, intersections and straightways environments and for different target rate requirements.
\begin{table}[H]
\caption{Percentage of locations above target rate before and after BS deployment manual modification.}
    \centering\label{bs}
    \begin{tabular}{|c|c|c|c|c|c|c|}
    \hline
    \textbf{Target rate}&\multicolumn{2}{|c|}{1 Gbps}& \multicolumn{2}{|c|}{500 Mbps}&\multicolumn{2}{|c|}{50 Mbps}\\ \hline
    \textbf{Deployment} & Before& After&Before& After&Before& After  \\ \hline
    \textbf{Corners} & 75.70 & 92.30 & 93.20 & 96.18 & 97.80 & 99.92 \\ \hline
    \textbf{Intersections} & 82.48  & 96.30 & 94.08 & 98.20 & 99.44 &99.98 \\ \hline  
    \textbf{Straightways} & 96.30  & 96.38 & 98.05 & 98.52 & 99.95 &99.99 \\ \hline  
    \end{tabular}
 \end{table}    

Regardless of the services requirements, we notice that the coverage enhances for the smart BS deployment compared to the pseudorandom realizations. The improvement is noticeable for corners and intersections environments and becomes negligible for the straightways scenarios. Considering the collective perception service, the percentage of users in coverage located in corners and intersections are 92.30$\%$ and 96.30$\%$, respectively, compared to the straightways where the coverage is 96.38$\%$. In fact, the smart BS deployment alleviates the effects of blockage and enables the users to connect to LOS BSs and hence it renders the corners and intersections environments less complicated like the straightways. Regarding the straightways, although the coverage is always better, the smart BS deployment has negligible effect on the rate and coverage for all the services requirements. This is explained by the fact that this scenario is already less complicated and most of the users are associated to LOS BSs. Consequently, the smart deployment will not be providing significant reliable connectivities to LOS serving BSs and the coverage will be roughly the same as the pseudorandom deployment. Note that this smart BS deployment applied in this work is manual and we expect that the coverage and rate can be further increased when the BS deployment is spatially optimized. We plan to defer the BS deployment optimization algorithm as a future extension for this work.

\section{Conclusion}
In this work, we provide a global simulation framework of a V2I urban street canyon scenario. Specifically, we constructed the rate map for analog and hybrid architectures for different combinations of the system parameters. For the analog architecture scenario, the results showed that the rate distribution improves with the carrier frequency, and the bandwidth. In addition, we illustrated the effects of blockage on the NLOS users who achieve low but non-zero rates because of the signal reflections from the buildings. The results also indicated that oversampled codebook improves the rate distribution in the grid. For the hybrid architecture, the results confirmed that the rate distribution enhances with improving the effective channel quantization. Also, we illustrate the trade-off between the rate and energy efficiency for partially-connected and fully-connected structures. In addition, we demonstrated that the rate map performance can be improved by considering the Zero-Forcing digital precoding to manage the multiuser interference. Finally, we finalized the work by collecting the rates statistics for single-user scenario to investigate the rate coverage for some services requirements in corners, intersections and straightways. We came up with the conclusion that the coverage is better in straightways than in corners and intersections as the blockers density is lower and the road geometry is less complicated. We further observed that the traffic intensity has negligible impact on the coverage and the smart BSs deployment signifcantly improves the spatial coverage compared to the pseudorandom BSs deployment. As future directions, we will elaborate an algorithm that optimizes the BSs deployment in order to maximize the spatial coverage and rate distribution. In addition, we intend to provide a tractable analysis using stochastic geometry approach to derive the joint coverage and rate in order to validate the rate map generated by the Ray-Tracing tool.

\bibliographystyle{IEEEtran}
\bibliography{main}

\begin{thebibliography}{10}
\providecommand{\url}[1]{#1}
\csname url@samestyle\endcsname
\providecommand{\newblock}{\relax}
\providecommand{\bibinfo}[2]{#2}
\providecommand{\BIBentrySTDinterwordspacing}{\spaceskip=0pt\relax}
\providecommand{\BIBentryALTinterwordstretchfactor}{4}
\providecommand{\BIBentryALTinterwordspacing}{\spaceskip=\fontdimen2\font plus
\BIBentryALTinterwordstretchfactor\fontdimen3\font minus
  \fontdimen4\font\relax}
\providecommand{\BIBforeignlanguage}[2]{{%
\expandafter\ifx\csname l@#1\endcsname\relax
\typeout{** WARNING: IEEEtran.bst: No hyphenation pattern has been}%
\typeout{** loaded for the language `#1'. Using the pattern for}%
\typeout{** the default language instead.}%
\else
\language=\csname l@#1\endcsname
\fi
#2}}
\providecommand{\BIBdecl}{\relax}
\BIBdecl

\bibitem{j3}
E.~{Balti} and M.~{Guizani}, ``Mixed rf/fso cooperative relaying systems with
  co-channel interference,'' \emph{IEEE Transactions on Communications},
  vol.~66, no.~9, pp. 4014--4027, 2018.

\bibitem{v1}
E.~Ohn-Bar and M.~M. Trivedi, ``{Looking at Humans in the Age of Self-Driving
  and Highly Automated Vehicles},'' \emph{IEEE Transactions on Intelligent
  Vehicles}, vol.~1, no.~1, pp. 90--104, March 2016.

\bibitem{v2}
J.~Pimentel and J.~Bastiaan, ``{Characterizing the Safety of Self-Driving
  Vehicles: A Fault Containment Protocol for Functionality Involving Vehicle
  Detection},'' in \emph{2018 IEEE International Conference on Vehicular
  Electronics and Safety (ICVES)}, Sep. 2018, pp. 1--7.

\bibitem{v3}
F.~Jimenez, J.~E. Naranjo, J.~J. Anaya, F.~Garcia, A.~Ponz, and J.~M. Armingol,
  ``{Advanced Driver Assistance System for Road Environments to Improve Safety
  and Efficiency},'' \emph{Transportation Research Procedia}, vol.~14, pp. 2245
  -- 2254, 2016, transport Research Arena TRA2016.

\bibitem{c1}
E.~{Balti}, M.~{Guizani}, B.~{Hamdaoui}, and Y.~{Maalej}, ``Partial relay
  selection for hybrid rf/fso systems with hardware impairments,'' in
  \emph{2016 IEEE Global Communications Conference (GLOBECOM)}, 2016, pp. 1--6.

\bibitem{c4}
Y.~{Maalej}, A.~{Abderrahim}, M.~{Guizani}, B.~{Hamdaoui}, and E.~{Balti},
  ``Advanced activity-aware multi-channel operations1609.4 in vanets for
  vehicular clouds,'' in \emph{2016 IEEE Global Communications Conference
  (GLOBECOM)}, 2016, pp. 1--6.

\bibitem{n1}
N.~{Mensi}, M.~{Guizani}, and A.~{Makhlouf}, ``Study of vehicular cloud during
  traffic congestion,'' in \emph{2016 4th International Conference on Control
  Engineering Information Technology (CEIT)}, 2016, pp. 1--6.

\bibitem{n2}
N.~{Mensi}, A.~{Makhlouf}, and M.~{Guizani}, ``Incentives for safe driving in
  vanet,'' in \emph{2016 4th International Conference on Control Engineering
  Information Technology (CEIT)}, 2016, pp. 1--6.

\bibitem{c2}
E.~{Balti}, M.~{Guizani}, and B.~{Hamdaoui}, ``Hybrid rayleigh and
  double-weibull over impaired rf/fso system with outdated csi,'' in \emph{2017
  IEEE International Conference on Communications (ICC)}, 2017, pp. 1--6.

\bibitem{map}
W.~Zheng, A.~Ali, N.~González-Prelcic, R.~W. Heath, A.~Klautau, and E.~M.
  Pari, ``5g v2x communication at millimeter wave: rate maps and use cases,''
  in \emph{2020 IEEE 91st Vehicular Technology Conference (VTC2020-Spring)},
  2020, pp. 1--5.

\bibitem{va}
J.~{Choi}, V.~{Va}, N.~{Gonzalez-Prelcic}, R.~{Daniels}, C.~R. {Bhat}, and
  R.~W. {Heath}, ``{Millimeter-Wave Vehicular Communication to Support Massive
  Automotive Sensing},'' \emph{IEEE Communications Magazine}, vol.~54, no.~12,
  pp. 160--167, December 2016.

\bibitem{3gpp2}
G.~W.~G. 1, ``Ls on prioritised use cases and requirements for consideration in
  rel-16 nr-v2x,'' R1-1809720, Tech. Rep., Aug 2018.

\bibitem{3gpp3}
G.~T. 22.886, ``Study on enhancement of 3gpp support for 5g v2x services
  (release 16),'' Technical Report V16.2.0, Tech. Rep., Dec 2018.

\bibitem{3gpp4}
5G-PPP, ``5g empowering vertical industries,'' 5GPPP White Paper, Tech. Rep.,
  Feb 2016.

\bibitem{3gpp5}
G.~T. 38.885, ``Nr: Study on vehicle-to-everything,'' Tech. Rep., Nov 2018.

\bibitem{3gpp6}
G.~T. 37.885, ``Methodology of new v2x use cases for lte and nr,'' Release 15,
  Tech. Rep., Dec 2018.

\bibitem{lte}
P.~{Belanovic}, D.~{Valerio}, A.~{Paier}, T.~{Zemen}, F.~{Ricciato}, and C.~F.
  {Mecklenbrauker}, ``{On Wireless Links for Vehicle-to-Infrastructure
  Communications},'' \emph{IEEE Transactions on Vehicular Technology}, vol.~59,
  no.~1, pp. 269--282, Jan 2010.

\bibitem{j4}
E.~{Balti} and B.~K. {Johnson}, ``Tractable approach to mmwaves cellular
  analysis with fso backhauling under feedback delay and hardware
  limitations,'' \emph{IEEE Transactions on Wireless Communications}, vol.~19,
  no.~1, pp. 410--422, 2020.

\bibitem{decade}
K.~{Sato} and M.~{Fujise}, ``{Propagation Measurements for Inter-Vehicle
  Communication in 76-GHz Band},'' in \emph{2006 6th International Conference
  on ITS Telecommunications}, June 2006, pp. 408--411.

\bibitem{surv}
V.~Va, T.~Shimizu, G.~Bansal, and R.~W. Heath, Jr., \emph{{Millimeter Wave
  Vehicular Communications: A Survey}}.\hskip 1em plus 0.5em minus 0.4em\relax
  Hanover, MA, USA: Now Publishers Inc., 2016.

\bibitem{tassi}
A.~{Tassi}, M.~{Egan}, R.~J. {Piechocki}, and A.~{Nix}, ``{Modeling and Design
  of Millimeter-Wave Networks for Highway Vehicular Communication},''
  \emph{IEEE Transactions on Vehicular Technology}, vol.~66, no.~12, pp.
  10\,676--10\,691, Dec 2017.

\bibitem{urban}
Y.~{Wang}, K.~{Venugopal}, A.~F. {Molisch}, and R.~W. {Heath}, ``{MmWave
  Vehicle-to-Infrastructure Communication: Analysis of Urban Microcellular
  Networks},'' \emph{IEEE Transactions on Vehicular Technology}, vol.~67,
  no.~8, pp. 7086--7100, Aug 2018.

\bibitem{inv}
V.~{Va}, J.~{Choi}, T.~{Shimizu}, G.~{Bansal}, and R.~W. {Heath}, ``{Inverse
  Multipath Fingerprinting for Millimeter Wave V2I Beam Alignment},''
  \emph{IEEE Transactions on Vehicular Technology}, vol.~67, no.~5, pp.
  4042--4058, May 2018.

\bibitem{pos}
V.~{Va}, T.~{Shimizu}, G.~{Bansal}, and R.~W. {Heath}, ``{Position-aided
  millimeter wave V2I beam alignment: A learning-to-rank approach},'' in
  \emph{2017 IEEE 28th Annual International Symposium on Personal, Indoor, and
  Mobile Radio Communications (PIMRC)}, Oct 2017, pp. 1--5.

\bibitem{prediction}
Y.~{Wang}, M.~{Narasimha}, and R.~W. {Heath}, ``{MmWave Beam Prediction with
  Situational Awareness: A Machine Learning Approach},'' in \emph{2018 IEEE
  19th International Workshop on Signal Processing Advances in Wireless
  Communications (SPAWC)}, June 2018, pp. 1--5.

\bibitem{c3}
E.~{Balti}, M.~{Guizani}, B.~{Hamdaoui}, and B.~{Khalfi}, ``Mixed rf/fso
  relaying systems with hardware impairments,'' in \emph{GLOBECOM 2017 - 2017
  IEEE Global Communications Conference}, 2017, pp. 1--6.

\bibitem{spatially}
A.~F. {Molisch}, A.~{Karttunen}, S.~{Hur}, J.~{Park}, and J.~{Zhang},
  ``{Spatially consistent pathloss modeling for millimeter-wave channels in
  urban environments},'' in \emph{2016 10th European Conference on Antennas and
  Propagation (EuCAP)}, April 2016, pp. 1--5.

\bibitem{cross}
Y.~{Wang}, K.~{Venugopal}, A.~F. {Molisch}, and R.~W. {Heath}, ``{Blockage and
  Coverage Analysis with MmWave Cross Street BSs Near Urban Intersections},''
  in \emph{2017 IEEE International Conference on Communications (ICC)}, May
  2017, pp. 1--6.

\bibitem{j1}
E.~{Balti} and M.~{Guizani}, ``Impact of non-linear high-power amplifiers on
  cooperative relaying systems,'' \emph{IEEE Transactions on Communications},
  vol.~65, no.~10, pp. 4163--4175, 2017.

\bibitem{j2}
E.~{Balti}, M.~{Guizani}, B.~{Hamdaoui}, and B.~{Khalfi}, ``Aggregate hardware
  impairments over mixed rf/fso relaying systems with outdated csi,''
  \emph{IEEE Transactions on Communications}, vol.~66, no.~3, pp. 1110--1123,
  2018.

\bibitem{ray}
\BIBentryALTinterwordspacing
``{Wireless InSite 3D Wireless Prediction Software}.'' [Online]. Available:
  \url{https://www.remcom.com/wireless-insite-em-propagation-software}
\BIBentrySTDinterwordspacing

\bibitem{ni}
\BIBentryALTinterwordspacing
``{mmWave: The Battle of the Bands}.'' [Online]. Available:
  \url{http://www.ni.com/en-us/innovations/white-papers/16/mmwave--the-battle-of-the-bands.html}
\BIBentrySTDinterwordspacing

\bibitem{anum}
A.~Ali, N.~{Gonz{\'{a}}lez Prelcic}, and R.~W. {Heath}, ``{Spatial Covariance
  Estimation for Millimeter Wave Hybrid Systems using Out-of-Band
  Information},'' \emph{CoRR}, vol. abs/1804.11204, 2018.

\bibitem{energy}
X.~{Yu}, J.~{Shen}, J.~{Zhang}, and K.~B. {Letaief}, ``{Alternating
  Minimization Algorithms for Hybrid Precoding in Millimeter Wave MIMO
  Systems},'' \emph{IEEE Journal of Selected Topics in Signal Processing},
  vol.~10, no.~3, pp. 485--500, April 2016.

\end{thebibliography}
\end{document}